\documentclass[preprint,showpacs,preprintnumbers,amsmath,amssymb]{revtex4}

\usepackage{graphicx}
\usepackage{dcolumn}
\usepackage{bm}

\begin{document}


\title{The Effects of Phase Separation in the Cuprate Superconductors\\}

\author{E. V. L. de Mello} 
\altaffiliation[]{evandro@if.uff.br}  
\author{E. S. Caixeiro}
\affiliation{%
Instituto de F\'{\i}sica, Universidade Federal Fluminense, Niter\'oi, RJ 24210-340, Brazil\\}%

\date{\today}

\begin{abstract}
 
 Phase separation has been observed by several different experiments
and it is believed to be closely related with the physics of cuprates but
its exactly role is not yet well known. We 
propose that the onset of pseudogap phenomenon or the
upper pseudogap temperature $T^*$ has its origin in a spontaneous phase separation 
transition at the temperature $T_{ps}=T^*$.
In order to perform quantitative calculations,
we use a Cahn-Hilliard (CH) differential equation 
originally proposed to the studies of alloys 
and on a spinodal decomposition mechanism. Solving numerically
the CH equation it is possible to follow the time evolution of a
coarse-grained order parameter which satisfies a 
Ginzburg-Landau free-energy functional commonly used to model 
superconductors. In this approach, we follow the 
process of charge segregation into two main equilibrium hole density branches and the
energy gap normally attributed to the upper pseudogap  arises 
as the free-energy potential barrier between these two
equilibrium densities below $T_{ps}$. This 
simulation provides  quantitative results 
in agreement with 
the observed stripe and granular pattern of segregation. Furthermore, with
a Bogoliubov-deGennes (BdG) local  superconducting 
critical temperature calculation for the lower pseudogap or the
onset of local superconductivity, it yields novel
interpretation of several non-conventional measurements
on cuprates.

\end{abstract}

\pacs{74.72.-h, 74.80.-g, 74.20.De, 02.70.Bf}
\maketitle

\section{Introduction}

 The existence of the pseudogap in all family of high-temperature superconductors
(HTSC) has been verified by several different experimental techniques
as discussed by many reviews\cite{TS,Tallon}.
As a consequence of many years of scientific effort, there is a solid 
consensus of its existence at least in the  underdoped regime. On
the other hand, there is currently no agreement on such basic facts
as to its nature and origin. After its discovery\cite{Alloul,Tallon},
it was realized that some experiments detected the pseudogap
temperature $T^*$ at very high values while others would
place it just above the critical temperature $T_c$. This  is probably 
because different probes are able to detect different properties but, the
fact is that this large discrepancy 
triggered a variety of different proposals. Just to mention
a few ideas and works; Emery et al.\cite{Emery} called the high $T^*$ as
$T_1^*$, the crossover temperature at which charge inhomogeneities
become well defined and the low $T^*$ as $T_2^*$ and associated it
with a spin gap and they both merged into $T_c$ at the slightly
overdoped region of the phase diagram. In their review Timusk and 
Statt\cite{TS}
also presented a similar phase diagram but they related the lower pseudogap
temperature to $T^*$ and the upper one also to a crossover temperature $T_1^*$. 
The lower and the higher $T^*$ were also considered as the opening
of a spin and a charge gap respectively\cite{MKZD}. The lower $T^*$
was also attributed to superconducting phase fluctuations\cite{EK}
and many different experiments claimed to have detected 
such fluctuations\cite{Corson,Bernhard,Xu,Wang,Meingast}. Thus the existence
of the two pseudogaps  in the cuprates has been compiled by several
works\cite{TS,Tallon,Markiewicz} as the result of many different data. 
In fact, analyzing the data from angle-resolved photoemission (ARPES) and angle-integrated
photoemission (AIPES), Ino et al.\cite{Ino} could distinguish not
two but three different energy scales.

 Another controversial point is whether the pseudogap and the superconducting
gap have the same origin or not. Tunneling spectroscopy\cite{Renner,Miyakawa,
Suzuki} seems to show that the gap evolves 
continuously from the superconducting into the normal phase without any 
anomaly, suggesting that the pseudogap and superconducting gaps have
the same origin. The common origin was also supported by some ARPES\cite{Harris}
and scanning tunneling spectroscopy (STM)\cite{Kugler} data. 
Muon spin rotating experiments\cite{Uemura} characterized  $T^*$ as the pair
formation line in agreement with the fluctuation theories of pre-formed 
superconducting pairs\cite{Emery,EK,Pieri}. These STM and ARPES
experiments have also measured the pseudogap in the overdoped region
in opposition to many others\cite{TS,Tallon,Uemura} which
the pseudogap temperature line appears to fall a little beyond 
the optimum doping value. On the other hand,
intrinsic (c-axis interplane) tunneling spectroscopy\cite{Krasnov00,Krasnov,Yurgens} 
led to results against a superconducting origin of the pseudogap what
was also confirmed by the same type of experiment 
in high magnetic field\cite{Krasnov}.
This conclusion, against the common origin of the pseudogap and
superconducting gap, is also shared by Tallon and Loram after the
analysis of data from many different experiments\cite{Tallon}.

 The above resumed paragraphs intended to show that, despite the enormous
experimental effort after all these years, 
there are still some basic open questions in this field.
These open questions motivated us to make the present novel work which
connects the large pseudogap  $T^*$ to the onset of phase separation.
There is now considerable evidence that the tendency toward phase
separation or intrinsic hole clustering formation is an universal 
feature of doped cuprates\cite{Muller,Castro,Statt,CEKO,Wang2}.
Phase separation in hole rich and hole poor regions was theoretically
predicted\cite{Zaanen} and has been observed
in the form of stripes\cite{Tranquada,Bianconi} and in the form of 
microscopic grains or mesoscopic segregation by STM 
measurements\cite{Pan,Davis}. Although the STM results has been questioned
as a surface phenomena which does  not reflect the nature of the
bulk electronic state\cite{Loram}, the inhomogeneities has also been seen
by neutron diffraction\cite{Tranquada,Bianconi,Buzin}  
which is essentially a  bulk type probe
in underdoped and optimally doped region of the $La_{2-x}Sr_xCuO_4$ 
phase diagram.
Another bulk-type measurement using nuclear quadrupole resonance  (NQR)\cite{Singer}
has observed an increase in the  hole density spatial variation 
of $La_{2-x}Sr_xCuO_4$ compounds  (with $0.04\le x\le 0.15$) 
as function of the temperature. Despite
these evidences, the majority of the theoretical approaches are based on the
assumption that the holes are homogeneously doped into $CuO$ planes, probably
due to the argument that, in principle, macroscopic phase separation 
is prevented  by the large Coulomb energy cost of concentrating 
doped holes into  small regions. On the other hand, the above cited references  are
just a few of the large number of works which have detected some type
of inhomogeneities in cuprates which seems to be intrinsic since it is
present even in the best single crystals\cite{CEKO}. 
There are also experimental evidences
for an intrinsic phase separation and cluster formation in many other
materials like, for instance,  manganites which are believed to be
another strong correlated electron materials\cite{AMoreo,Dagotto,Dagotto1} and
on rutheno-cuprates superconductors\cite{Chu}. In fact, it has been argued
that phase separation might be stronger in manganites\cite{AMoreo} than
in cuprates.

In this article we  develop a novel approach to this issue as
we apply to the large pseudogap  $T^*$ the theory
of phase-ordering dynamics, that is, the growth of domain coarsening
when a system is quenched from the homogeneous phase into an 
inhomogeneous phase\cite{Bray}. 
This phenomenon is also known as {\it spinodal decomposition}.  
One of the leading
models devised for the theoretical study of this phenomenon for a
conservative order parameter is
based on the Cahn-Hilliard formulation\cite{CH}.  The Cahn-Hilliard
(CH) theory was originally proposed to model the quenching of binary
alloys through the critical temperature but it has subsequently
been adopted to model many other physical systems which go through a
similar phase separation\cite{Bray,CH,Eyre}.  We show how the CH equation is 
derived from a typical Ginzburg-Landau (GL) free energy for a typical 
(conserved) order parameter, which is easily related with the 
density of holes, using an equation for the conservation of the
order parameter current. The CH equation is solved numerically 
by adopting a very efficient method (compared with usual first order
Euler methods) semi-implicit (in time) finite difference scheme proposed by 
Eyre\cite{Eyre}. The numerical details have been analyzed elsewhere\cite{Otton}. 

The main purpose to  solve the CH equation
for the hole density field and take the large pseudogap temperature $T^*$
as the phase separation temperature $T_{ps}$ 
is that  we can make quantitative calculations and  get some insights on various 
HTSC non-conventional features: as the temperature goes down below
$T_{ps}$, the distribution of hole 
density for a given compound evolves smoothly from an initially random variation taken as 
a  Gaussian distribution around an average density $p$,
since a purely uniform distribution does not segregate, into
a kind of bimodal distribution. These simulations are used to
demonstrate the charge inhomogeneity and 
the stripe pattern formation in a square lattice as
shown below. The pseudogap energy $E_g$ or the large pseudogap 
temperature $T^*$ arises naturally as the GL potential barrier
between the two equilibrium density phases, changes smoothly
as the temperature decreases  and reaches the maximum phase
separation near zero temperature. If $T_{ps}$ vanishes at a critical
average hole density $p_c\approx 0.2$ as generally accepted\cite{TS,Uemura,Tallon},
that means that all the compounds with average $p\le p_c$ may undergo a phase
separation and evolves continuously into a complete separation characterized
by a bimodal distribution   with two major
equilibrium densities ($p_+$ and $p_-$). For underdoped samples 
the phase separation is more pronounced, 
since $T_{ps}$ is very large for these compounds. 
The difference between $p_+$ and $p_-$
should decrease for compounds with increasing average hole density $p$ and
the sharp peaks evolves into rounded peaks near
$p_c$. This provides an explanation for the neutron diffraction
data on the $Cu-O$ bond length distribution\cite{Buzin} and the 
observation of charge and spin separation
into stripe phases. On the other hand, the increase 
of the inhomogeneity (variation in $p$) as the temperature 
is decreased for a given sample was observed by
the NQR experiments\cite{Singer}, in agreement with the CH theory
of the spinodal decomposition. On the other hand,
these local differences in the charge distribution generate 
local microscopic (or mesoscopic) regions with different superconducting transition
temperatures. The onset of local superconductivity may be identified 
as the lower pseudogap temperature or the temperature where the superconducting
pairs start to appears. This second pseudogap has also been interpreted
as the mean field temperature $T^{MF}$ by Emery and Kivelson\cite{Emery,EK}. 
As the temperatures goes down between this lower 
$T^*$ and $T_c$ more superconducting regions or 
superconducting droplets appear, they grow in size and quantity
and they percolate at $T_c$. 
The appearance of these
superconducting  droplets above $T_c$ is in agreement and it is the only possible
explanation of various measurements made in the normal phase of different 
materials like the Nernst effect\cite{Xu,Wang} and the precursor 
diamagnetism\cite{Iguchi,Attilio1,Attilio2,Jorge}.
In this scenario,  superconducting phase coherence is achieved only at 
$T_c$ which is the temperature that $\approx60\%$ 
($\approx$ the percolating limit) of the sample 
volume is in the superconducting phase as has been proposed by 
several different works\cite{OWK,Mihailovic,Mello03}. In the following
sections we discuss the phase separation mechanism, we present
the results of some simulation and the implications
to HTSC properties in detail.

 As mentioned above, the process of phase separation in HTSC is well 
documented but, concerning the mechanism of phase separation there are
not  many conclusive studies. One possibility  for this 
mechanism arises from the  measurements by 
nuclear magnetic resonance (NMR)\cite{Statt}, which 
has determined the high mobility of the oxygen interstitial in 
$La_2CuO_{4-\delta}$ compounds. Therefore, it is possible that the dopant atoms
cluster themselves  to minimize the local energy an this would be
a possible explanation for the whole process. This is just a general idea
based on the NMR results\cite{Statt} but the mechanism of clustering is
an interesting subject that merits more attention in the future.

To avoid confusion in the notation, we will adopt  $T_{ps}(p)$ for the large
pseudogap temperature of a compound with average hole doping $p$ and  $T^*(p)$ for the lower 
pseudogap temperature. When we refer to a given sample and not to 
a family of compounds, to 
simplify the notation, we may just use $T_{ps}$ and $T^*$.

\section{The CH Approach to Phase Separation}

The CH theory was developed to the binary allows and one may question its
application to a strongly correlated  system as HTSC. However the
clustering process in hole doped HTSC is  very subtle. As we can draw from
the stripe phases, the antiferromagnetic insulating phase has nearly zero
holes per copper atom and the charged phase has less than 0.25 holes
per copper atom, and in some cases 0.125. Thus, double hole
occupancy does not occur in either phases,
which is in agreement with a large on-site coulomb repulsion used
in almost all Hamiltonian models for HTSC as in Eq.(\ref{Hext}) below. 
Therefore, we believe that the use of the CH theory to hole
doped HTSC is justified.

\begin{figure}[!ht]
\includegraphics[height=7cm]{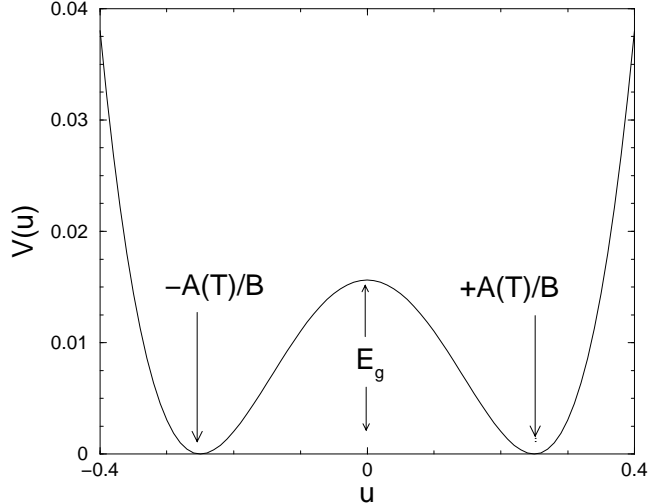}
\caption{ 
The typical potential used in the density of free energy which gives rise to
phase separation as function of the order parameter $u$. 
Notice that the two minima at $u_{\pm}$ yield the two equilibrium 
densities $p_{\pm}=u_{\pm}+p$ and the energy barrier  between them  $E_g$ depend on 
the temperature difference $T_{ps}-T$.}
\label{Vu}
\end{figure}

As an initial condition,
let's suppose that a typical HTSC has, above $T_{ps}$, a Gaussian distribution 
of local densities around an average hole density $p$  as can be 
direct inferred from the  STM experiments\cite{Pan,Davis}. Pan 
et al\cite{Pan} have measured a spread of $\Delta p\approx 0.08$ 
holes/Cu for an optimally doped compound which will be adopted as an 
initial condition in our calculations. This Gaussian distribution  around
the average hole density $p$ is the starting point
at temperatures above and near the phase separation temperature $T_{ps}$ 
and each local hole density $p(\vec x)$ inside the sample oscillates around the 
compound average $p$. In this way, we
can define the order parameter $u(\vec x)\equiv p(\vec x)-p$ and
$u(\vec x)=0$ above and at $T_{ps}$, as expected. Then
the typical GL functional for the free energy density in terms
of such order parameter is 
\begin{eqnarray}
f= {{{1\over2}\varepsilon^2 |\nabla u|^2 +V(u)}}
\label{GL}
\end{eqnarray}
where the potential $V(u)= A^2(T)u^2/2+B^2u^4/4+...$, 
$A^2(T)=\alpha(T-T_{ps})$ and B is a constant.
Notice that near and below $T_{ps}$ and/or for small values of $\varepsilon$, the
gradient term can be neglected and we get the two minima of $f$ at
the equilibrium values $u(\vec x)=\pm A/B=\pm \sqrt(\alpha(T_{ps}-T))/B$. 
This can be easily seen if we write $V(u)=B^2(u^2-A^2/B^2)^2$. On Fig.(\ref{Vu})
we show the important characteristics of such potential: As the temperatures
go down away from $T_{ps}$, the two equilibrium order parameter 
(or densities) go further 
apart from one another and the energy barrier
between the two equilibrium phases $E_g$ also increases. $E_g=A^4(T)/B$
which is proportional to $(T_{ps}-T)^2$.

Bray\cite{Bray}
pointed out that one can explore the fact that the type of order
parameter used above, as the two types of atoms of a given alloy,
is conserved and the CH equation can be written in the
form of a continuity equation, $\partial_tu=-{\bf \nabla.J}$,
with the current ${\bf J}=M{\bf \nabla}(\delta f/ \delta u)$,
where $M$ is the mobility or the transport coefficient. It is
probably the same for each family of HTSC compounds because of
the universal character of their phase diagram.
Therefore we may write the CH equation as following,
\begin{eqnarray}
\frac{\partial u}{\partial t} = -M\nabla^2(\varepsilon^2\nabla^2u
+ A^2(T)u-B^2u^3).
\label{EqCH}
\end{eqnarray}  

This equation is solved
with the so-called flux-conserving boundary conditions, 
${\bf \nabla}u.\vec n|_{\vec x\in\partial\Omega}=
(\nabla^3u).\vec n|_{\vec x\in\partial\Omega}=0$
where $\vec n$ is the outward normal vector on the boundary of the domain
$\Omega$ which we represent by $\partial\Omega$, it is possible
to show the time conservation of the total mass $M_t$ and that the total free 
energy can only decrease (dissipate) or being stable\cite{Eyre,Otton}.
Therefore a time stepping finite difference scheme is defined to be {\it gradient
stable} only if the free energy is non-increasing and 
gradient stability is regarded as the best stability criterion for
finite difference numerical solutions of such non-linear partial 
differential equation as the
CH equation\cite{Otton}.

As it
has already been pointed out\cite{Eyre,Otton},
both the $\nabla^4$ and the
non-linear term make the CH equation very stiff and it is
difficult to solve it numerically. The non-linear term in principle, forbids the
use of common Fast Fourier Transform (FFT) methods and brings the
additional problem that the usual
stability analysis like von Neumann criteria cannot be used. These difficulties
make most of the finite
difference schemes to use time steps of many order of magnitude
smaller than $\Delta x$ and consequently, it is numerical expensive to reach
the time scales where the interesting dynamics occur. To solve these difficulties
Eyre proposed a semi-implicit method in time that is unconditional
gradient stable when the  $V(u)$ can be divided in two parts:
$V(u)=V_c(u)+V_e(u)$ where $V_c$ is called contractive and
$V_e$ is called expansive\cite{Eyre}. Thus, we adopt here his method
taking $V_e$ as the quadratic
term and $V_c$ as the forth order one. Then we finally  obtain the
proposed finite difference scheme for the CH equation which
is linearized in time (we have absorbed $M$ into the time step), 
namely\cite{Otton},
\begin{eqnarray}
&&U^{n+1}_{ijk}+\Delta t(\varepsilon^2\nabla^4U^{n+1}_{ijk}+
B^2\nabla^2(U^{n}_{ijk})^2U^{n+1}_{ijk}) \nonumber \\
&&= U^n_{ijk}-\Delta t A^2(T)\nabla^2U^n_{ijk}.
\label{EqCHEt}
\end{eqnarray}

We have studied the stability conditions of this equation in one, two
and three dimensions\cite{Otton}.
In the next section we present the results for two and three dimensions 
applied to the problem of phase separation in a HTSC plane
of CuO. Although we calculate the local order parameter $u(\vec x)$ of
a sample with average hole density p,
we are interested and will preferably refer to the local hole
density $p(\vec x)=u(\vec x)+p$.

\section{The Results of the Simulations}

As mentioned in the introduction,  there is a consensus 
from several different experiments\cite{TS,Tallon} that the pseudogap temperature
$T^*(p)$ initiate at average hole doping $p\approx0.05$ at $T\approx800$K and
falls to zero temperature at a critical doping $p_c\approx0.2$. 
This is best illustrated by Fig.(11) from the review work of Tallon et al\cite{Tallon}
with many different data, which we reproduce here for convenience.
\begin{figure}[!ht]
\begin{center}
\includegraphics[height=7cm]{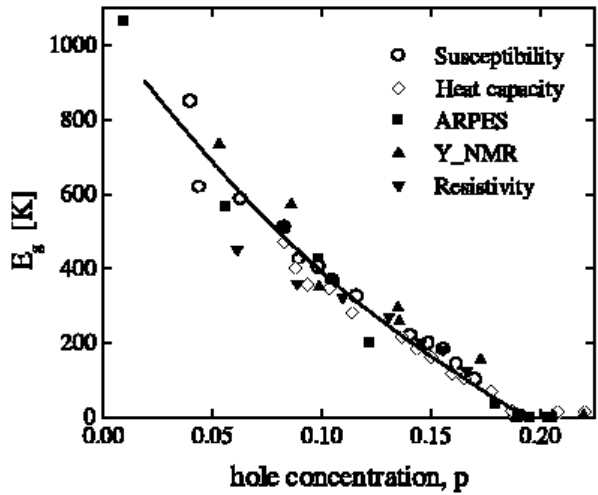}
\caption{ Fig.11 from Tallon et al\cite{Tallon} showing the p-dependence
of the pseudogap energy $E_g$ or $T^*$ determined from susceptibility, heat
capacity, ARPES, $^{89}$Y$\_$NMR and resistivity as displayed in the legends.}
\label{Tallonfig}
\end{center}
\end{figure}

Initially, that is above $T_{ps}(p)$,  the system has a homogeneous 
distribution of charge with
very small variations around $p$, which is described by a very narrow Gaussian-type
distribution. When the
temperature goes down through  $T_{ps}$ the sample with average hole
density $p$ starts to phase 
separate and the original Gaussian distribution of holes changes continuously
into a bimodal type distribution. 
For underdoped samples with large $T_{ps}$, the mobility $M$ is high 
which favors a rapid phase separation into
two main hole densities $p_-$ and $p_+$, while the compounds near
the critical doping $p_c$ may not undergo a complete 
phase separation. Near the $T_{ps}$, the difference
between  $p_-$ and $p_+$ is very small and increases as the 
temperature goes away from  $T_{ps}$. However if the system is quenched very rapidly
the phase separation may even not occur, because it depends on
the mobility which is essentially the phase separation time scale\cite{Bray,Otton}. 
For $p\ge p_c$ there is no phase separation and 
the charge distribution remains a Gaussian like. For $p\le p_c$,
the transformation from a homogeneous
phase to one with different densities and with sites at different
environment is seemed by many different measurements: 
by local measurements like, for instance, the 
Y$\_$NMR, by transport measurements like the 
resistivity since the charges must overcome the potential
barrier $E_g$ between the two equilibrium regions (see Fig.(\ref{Vu}))
and by susceptibility due to the appearance of  antiferromagnetic regions 
with low hole density specially at
the low average doping compounds.
Notice that the coefficient $A(T)=\sqrt(\alpha(T_{ps}-T))$ 
changes smoothly as the
temperature goes down away from $T_{ps}$ and therefore the charge distribution
in a given compound depends strongly on the temperature $T$,
on the details of sample synthesis and annealing procedures
and, due to the mobility,  on  
how the system is quenched through $T_{ps}$. 
This is probably the explanation
to the different results reported in the literature on 
many HTSC compounds.

\begin{figure}[!ht]
\includegraphics[height=7cm]{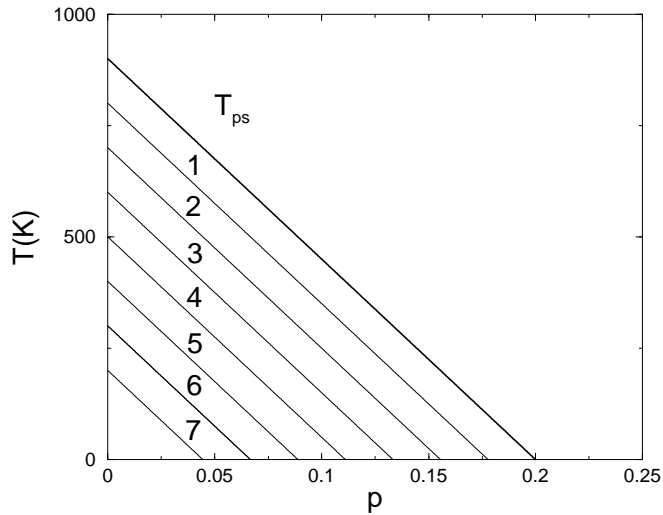}
\caption{Illustration of the phase separation regions. The thick line
represents $T_{ps}$ or $E_g$ from  Fig.(\ref{Tallonfig}) 
approximated by
a straight line. The numbered regions are equidistant from $T_{ps}$ and
are characterized by their single values of $(T_{ps}-T)$ which is 
proportional to the equilibrium densities $p_-$ and $p_+$.}
\label{Regions}
\end{figure}

Assuming that the curve proposed by Tallon and Loram\cite{Tallon} reproduced here
in the Fig.(\ref{Tallonfig}) is the $T_{ps}$ line, the regions below  are 
characterized by their temperature distance from this temperature. 
The regions in the bottom like 5 to 7, as illustrated
in Fig.(\ref{Regions}), are region with very strong phase
separation while regions near $T_{ps}$ like 1 to 3 the phase 
separation is weak. This is because $u_{\pm}=\pm(A/B)=\pm\sqrt(\alpha(T_{ps}-T)/B)$
and these regions are characterized by their values of $(T_{ps}-T)$. 
Thus, in region one, the difference between $p_-$ and $p_+$ is very
small and increases as the temperature goes below the $T_{ps}$ line. Accordingly, 
the energy gap $E_g=E_g(T)$ is a  varying function of $T$ and goes to 
zero near $T_{ps}$.  At zero temperature, compounds with
$p\le0.1$ may be strongly separated in a insulator phase ($p_-\approx 0$) and
in a metallic phase with  $p_+\ge 0.2$. Compounds with $0.1\le p \le 0.16$ the phase separation
is partial and for $0.16\le p \le 0.2$ the original Gaussian is distorted
with an increase in the hole density at the low and high tail.

We have performed calculations in all regions below the phase separation
line increasing the value of the $A$ coefficient simulating the
the temperature difference $(T_{ps}-T)$. Different
initial conditions were tested to check convergence after thousands of time
steps. One of the trial starting  initial
condition was, for instance, $u(t=0)=\varepsilon\times sin(x)sin(y)$.

\begin{figure}[!ht]
\includegraphics[height=6cm]{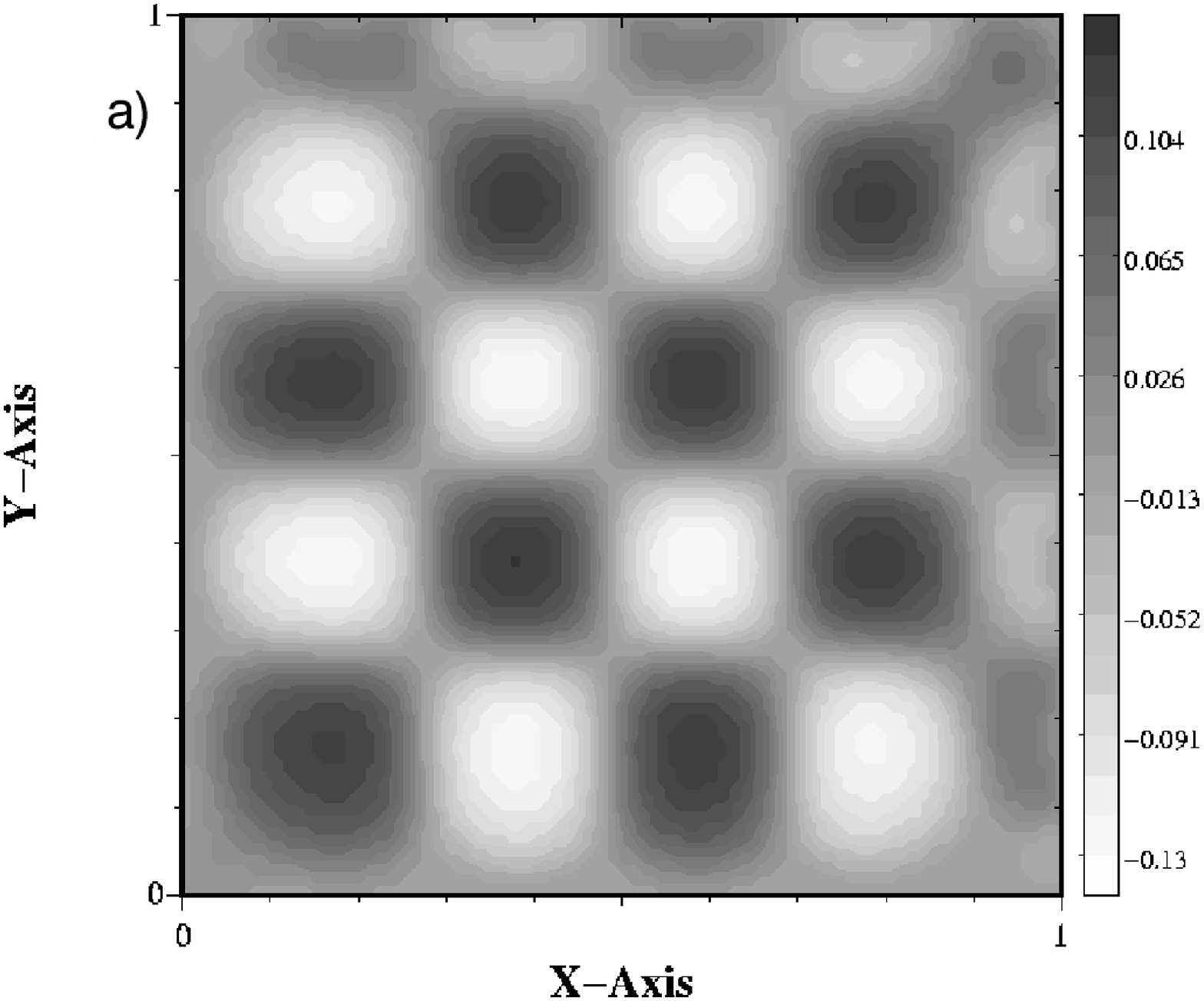}
\includegraphics[height=6cm]{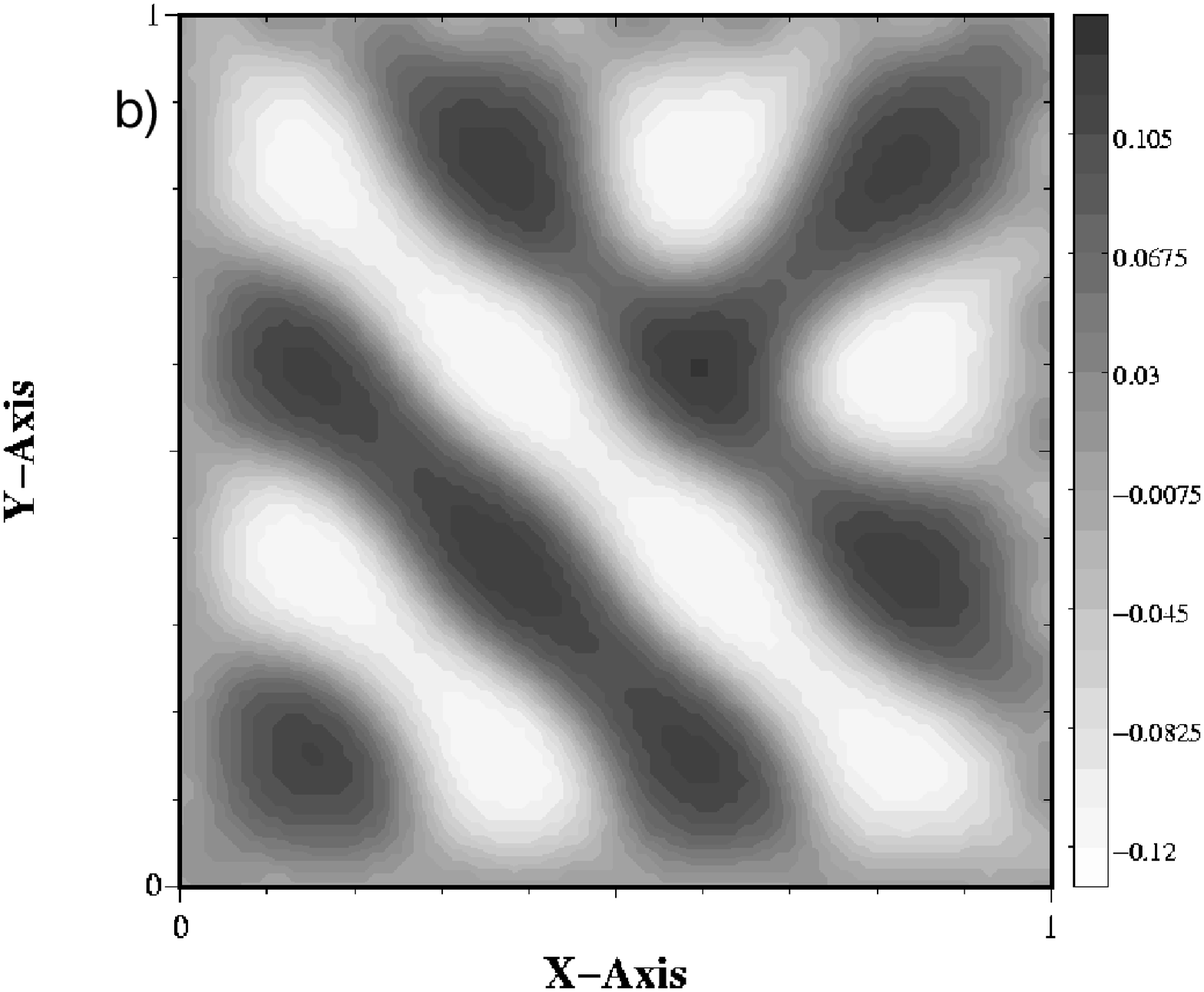}
\includegraphics[height=6cm]{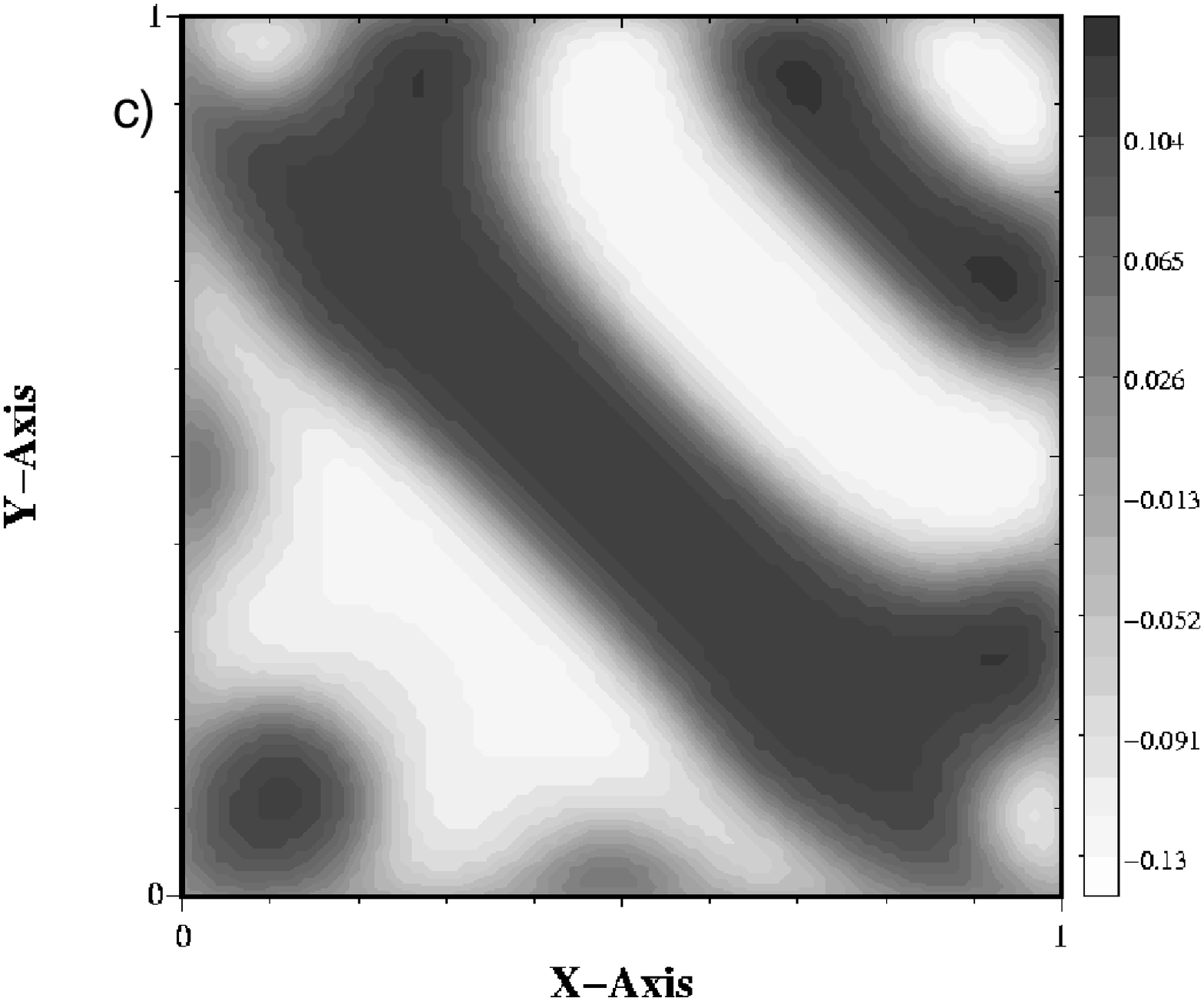}
\includegraphics[height=6cm]{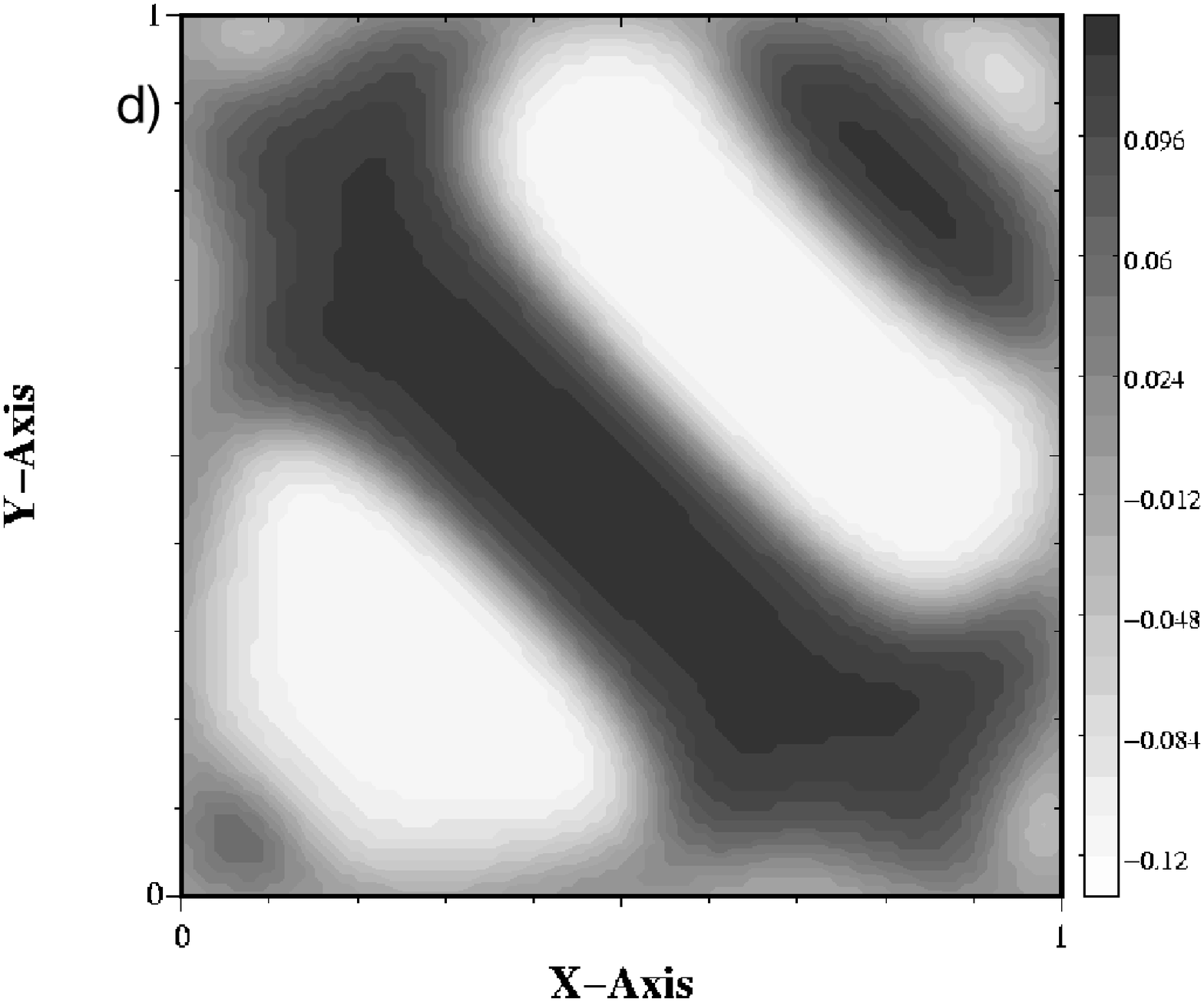}
\caption{ The  process of phase separation with the time. On the
top panel we plot the order parameter map at  the times  $t=2000$(a)
which displays an (enhanced)  reminiscent pattern from the initial conditions, 
At $t=5000$(b) the phase separation process has already
started and on the c) panel at the $t=10000$ and $t=25000$(d).}
\label{timevol}
\end{figure}

\begin{figure}[!ht]
\includegraphics[height=7cm]{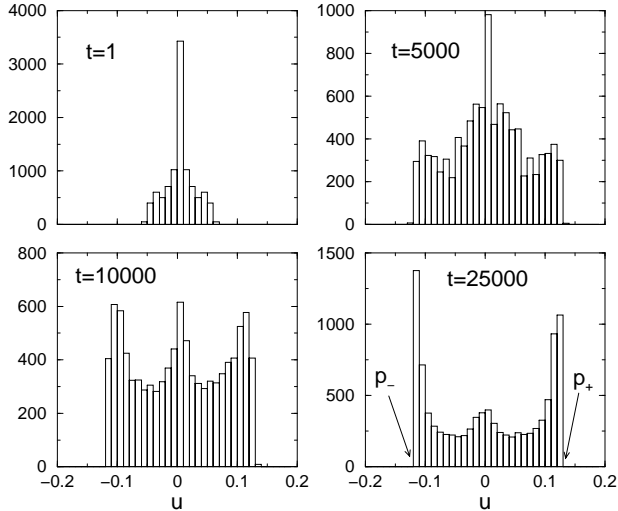}
\caption{The  evolution of the local densities of order parameter $u(\vec x)$ with the
time in our simulations. We can see the tendency toward  sharps  bimodal distributions
at the density equilibrium values ($u_-$ and $u_+$).  }
\label{histoevol}
\end{figure}

In Fig.(\ref{timevol}) we show the results of the simulations on a $100\times100$ square grid. 
In these simulations we used $A/B=0.125$ and $\varepsilon=0.05$ 
which represents a phase separation
in region 4 of Fig.(\ref{Regions}) because it is a region where phase
separation is neither minimal as in region 1 nor maximal as region 7. 
The simulation describes the time
evolution of a homogeneous initial condition given above and 
represented by a very  sharp Gaussian around the average p value 
shown in Fig.(\ref{histoevol}(a)). 
Fig.(\ref{timevol}) shows very clearly  the phase separation
process. 

The phase separation time evolution is also well illustrated by
displaying the histogram of how the order parameter evolves in time.
In the Fig.(\ref{histoevol}) we show the time evolution of a typical
simulation with the same parameters of Fig.(\ref{histoevol}): $t=1$ represents 
the initial condition with the hole
density $p(\vec x)$ centered around an average value $p=0.125$, $t=5000$
represents 5000 time steps in our simulations and so on. It is very
interesting the shape of these histogram and their evolution from
a initial centered Gaussian to a bimodal distribution. It is very important
to emphasize that the distribution after a certain time steps is
independent of the initial condition. In practice, if the mobility
would be very large and if $\varepsilon$ is very small, the system 
would evolve to two delta functions at $p_{\pm}$. 

\begin{figure}[!ht]
\includegraphics[height=7cm]{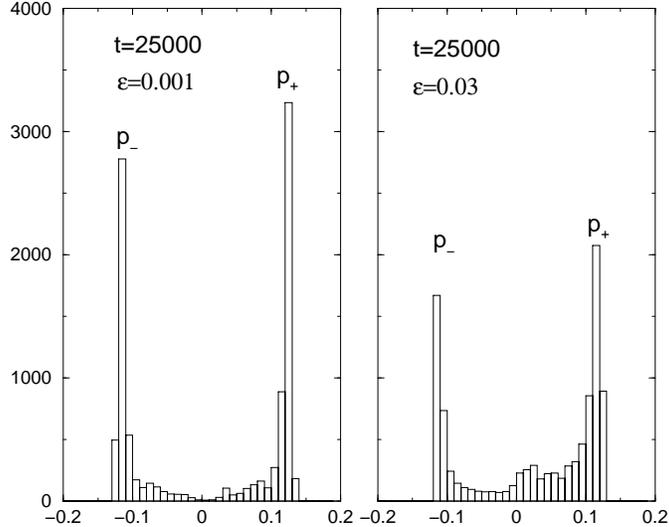}
\caption{The  evolution of the local densities of order
parameter probability with the
gradient constant $\varepsilon$. We can see the tendency toward sharp bimodal formation
at the values $p_{\pm}$ as $\varepsilon$ decreases.  }
\label{histepsi}
\end{figure}

To study the effect of the gradient term in the GL free energy of
Eq.(\ref{GL}) we have also performed simulations with different
values of $\varepsilon$. We have
tested $\varepsilon=0.01, 0.03$ and $0.05$. The results are shown 
in Fig.(\ref{histepsi}) and we can see
that indeed the order parameter distribution approaches a delta function
as $\varepsilon$ decreases.

\begin{figure}[!ht]
\includegraphics[height=9cm]{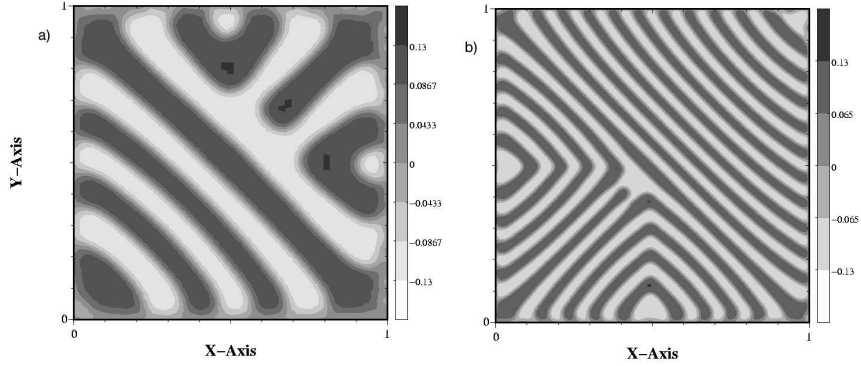}
\caption{ The mapping of the order parameter in the  process of phase separation in lattices with
different sizes. We display here the order parameter in the $200\times200$ (a)
and for  500$\times$500 (b) lattice. The parameter are the same and
therefore it is to be compared with the results displayed in Fig.4d above for the
$100\times100$ lattice.}
\label{size}
\end{figure}

Phase separation always occur when we start with a small variation 
around an average value but the final pattern is strongly
dependent on the size of the system. In order to study such effect
we have also done, together with the $100\times100$ lattice,
calculation with the $200\times200$ and $500\times500$ square grid. At
the Fig.(\ref{size}), 
we show the results  of mapping the order parameter in a surface 
with the same values of parameter used above. It is
very interesting, in the context of HTSC, to observe that smaller lattices
display granular pattern and there is a clear increase in the formation
of stripes pattern  as the size of the lattice is enlarged. It is a matter of
fact that the largest HTSC single crystals are those of the $La_{1-x}Sr_xCu0_2$
family which are more suitable for neutron diffraction studies and it is exactly in
this family which the stripe phases were measured\cite{Tranquada,Bianconi}. As
conjectured by A.Moreo et al\cite{AMoreo}, it is likely that the same conclusion
may be applied to the manganites.

\begin{figure}[!ht]
\includegraphics[height=9cm]{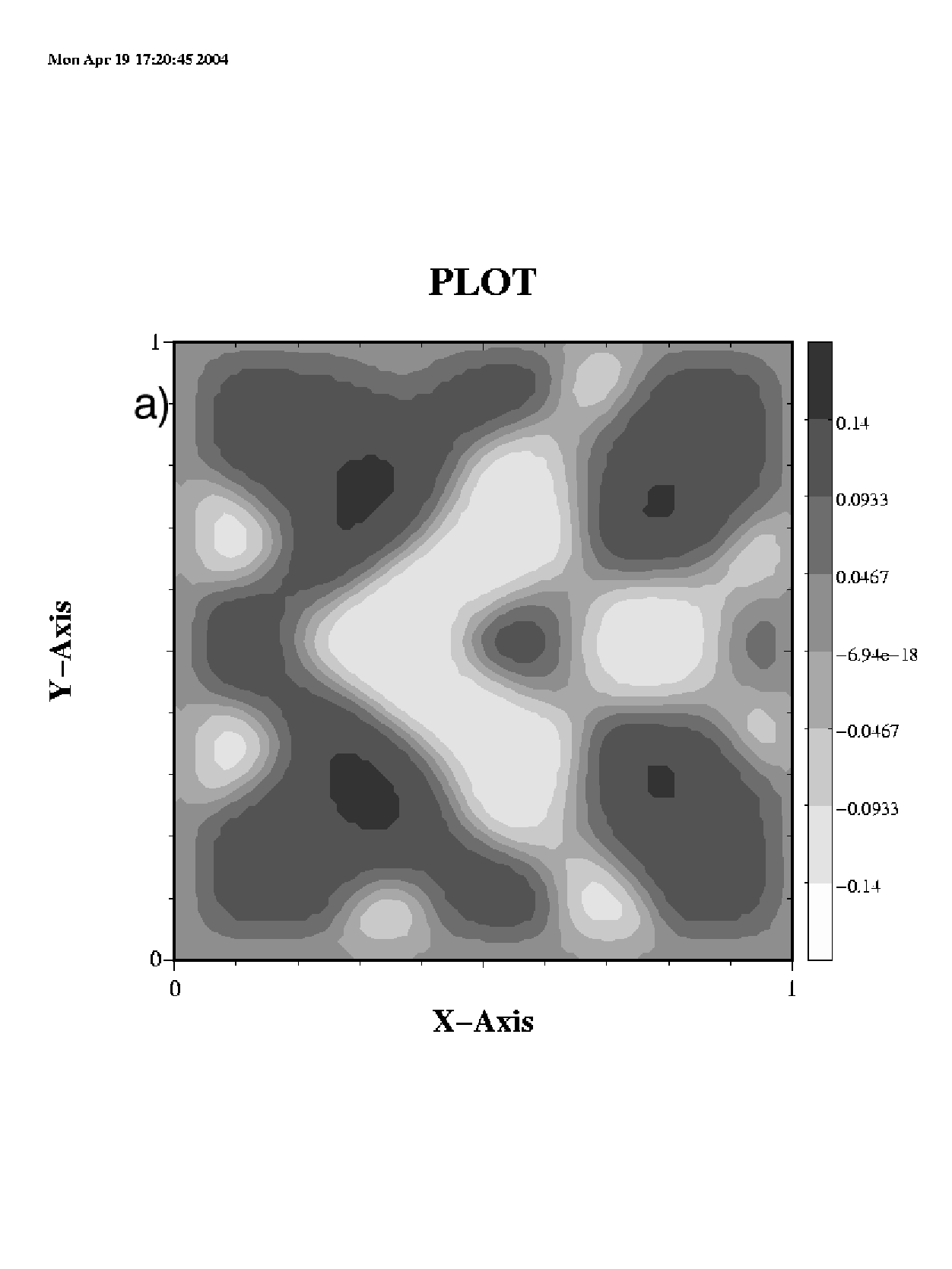}
\includegraphics[height=9cm]{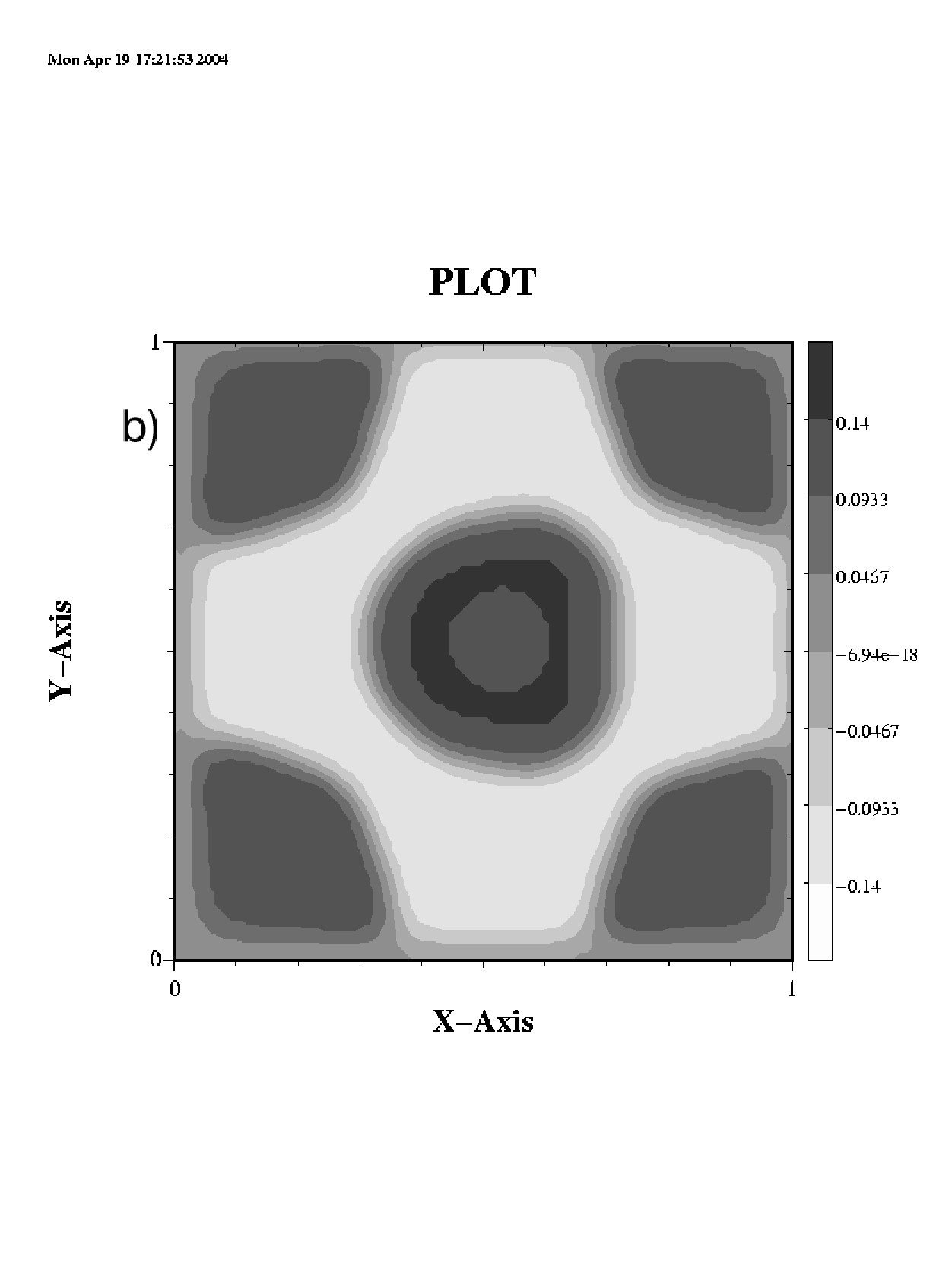}
\caption{The 3D mapping process of phase separation at t=50000 steps .On the
a) panel we plot the order parameter at the center $z=50$ of a $100\times100\times100$
lattice. At the b) panel we show the order  parameter near the top surface
($z=100$)}
\label{3D}
\end{figure}

 Notice how the stripe structure develops in the plane interior and
as they end at the borders they display a granular type
pattern similar to those found in STM\cite{Xu,Pan,Davis}. In
order the check this we have also performed simulations in 3D. The
results does not differ appreciably from the 2D case. At the Fig.(\ref{3D})
we show cuts in a three dimension $100\times100\times100$ lattice at the middle
plane ($z=50$) and near the top surface ($z=100$).

\section{The Local Gap}

We have shown that that below the $T_{ps}$ a phase separation develops
creating a variable density of holes at very
small mesoscopic scale. Therefore, it is important to perform a  local
superconducting gap calculation, taking into account this charge
inhomogeneity, in order to understand its effect on the  the
superconductivity phase and specifically, how such phase
is built in this inhomogeneous
environment. The appropriate way to do this calculation, in
a system without spatial invariance, is through the Bogoliubov-deGennes
(BdG) mean-field theory\cite{Soininen,Berlinsky,Franz,Ghosal,Ghosal2}. We start with
the extended Hubbard Hamiltonian

\begin{equation}
H=-\sum_{\ll ij\gg \sigma}t_{ij}c_{i\sigma}^\dag c_{j\sigma}
+\sum_{i\sigma}(V_i^{imp}-\mu)n_{i\sigma}
+U\sum_{i}n_{i\uparrow}n_{i\downarrow}+{V\over 2}\sum_{\langle ij\rangle \sigma
\sigma^{\prime}}n_{i\sigma}n_{j\sigma^{\prime}}
\label{Hext}
\end{equation}
where $c_{i\sigma}^\dag (c_{i\sigma})$ is the usual fermionic creation (annihilation)
operators at site ${\bf x}_i$,
spin $\sigma \lbrace\uparrow\downarrow\rbrace$, and
$n_{i\sigma} =  c_{i\sigma}^\dag c_{i\sigma}$.
$t_{ij}$ is the  hopping between site $i$ and $j$,
$U$ is the on-site and  $V$ is the nearest neighbor interaction.
$\mu$ is the chemical potential and  $V_i^{imp}$ is a random potential
which controls the strength of the disorder and introduces the
inhomogeneous Hartree shift\cite{Ghosal2}.

Using a mean-field decomposition approach, one can define the
pairing amplitudes\cite{Franz,Ghosal2},
$
\Delta_{\bf \delta}({\bf x}_i)=V \langle  c_{i\downarrow}c_{i+\delta\uparrow} \rangle
$
and
$\Delta_U({\bf x}_i)=U\langle c_{i\downarrow}c_{i\uparrow} \rangle$,
which yields an effective Hamiltonian
\begin{eqnarray}
H_{eff}
=-\sum_{i\delta\sigma}t_{i,i+\delta}c_{i\sigma}^\dag c_{i+\delta\sigma}
+\sum_{i\sigma}(V_i^{imp}-\tilde \mu_i)n_{i\sigma} \nonumber \\
+\sum_{i\delta }[\Delta_{\bf \delta}^*({\bf x}_i) c_{i\downarrow}c_{i+\delta\uparrow}
+ \Delta_{\bf \delta}({\bf x}_i) c_{i+\delta\uparrow}^\dag c_{i\downarrow}^\dag]
+ \sum_{i}[\Delta_U({\bf x}_i)c_{i\uparrow}^\dag c_{i\downarrow}^\dag
+\Delta_U^*({\bf x}_i)c_{i\downarrow}c_{i\uparrow}].
\label{Heff}
\end{eqnarray}

In this expression ${\bf \delta}$ represents the nearest neighbor
vectors and
$
\tilde \mu_i=\mu -{U\over 2}\langle n_i\rangle
$
is the  Hartree shift with the local eletronic density
$
\langle n_i\rangle=\sum_{\sigma}\langle n_{i\sigma}\rangle.
$ 
The hole density is 
$
p({\bf x}_i)=1-\langle n_i\rangle
$.
The $H_{eff}$ is diagonalized by the BdG transformation
\begin{eqnarray}
c_{i\uparrow}=&
\sum_n[\gamma_{n\uparrow}u_n({\bf x}_i)-\gamma_{n\downarrow}^{\dag}v_n^*({\bf x}_i)],\nonumber \\
c_{i\downarrow}=&
\sum_n[\gamma_{n\downarrow}u_n({\bf x}_i)+\gamma_{n\uparrow}^{\dag}v_n^*({\bf x}_i)]
\label{4v2}
\end{eqnarray}
where $\gamma_{n\sigma}$ and $\gamma_{n\sigma}^{\dag}$ are quasiparticle
operators associated with the excitation energies ($E_n\ge0$).
$u_n({\bf x}_i)$ and $v_n({\bf x}_i)$ are normalized amplitudes for
each ${\bf x}_i$. Therefore the BdG equations are 

\begin{equation}
\begin{pmatrix} K         &      \Delta  \cr\cr
           \Delta^*    &       -K^* 
\end{pmatrix}
\begin{pmatrix} u_n({\bf x_i})      \cr\cr
                v_n({\bf x_i})    
\end{pmatrix}=E_n
\begin{pmatrix} u_n({\bf x_i})       \cr\cr
                 v_n({\bf x_i})     
\end{pmatrix}
\label{matrix}
\end{equation}
with
\begin{eqnarray}
Ku_n({\bf x_i})&=&-\sum_{\delta}t_{i,i+\delta}u_n({\bf x}_i+{\delta})
+(V_i^{imp}-\tilde \mu_i)u_n({\bf x}_i)
\nonumber \\
\Delta u_n({\bf x_i})&=&\sum_{\delta}\Delta_{\delta}({\bf x}_i)u_n({\bf x}_i+{\delta})
+\Delta_U({\bf x}_i)u_n({\bf x}_i),
\end{eqnarray}
and similar  equations for $v_n({\bf x_i})$. These equations give the amplitudes
$(u_n({\bf x_i}),v_n({\bf x_i}))$, and the eigenergies $E_n$. The BdG equations
are solved self-consistently together with the pairing amplitude\cite{Berlinsky,Franz}

\begin{eqnarray}
\Delta_U({\bf x}_i)&=&-U\sum_{n}u_n({\bf x}_i)v_n^*({\bf x}_i)\tanh{E_n\over 2k_BT} ,
\label{DeltaU}
\end{eqnarray}

\begin{eqnarray}
\Delta_{\delta}({\bf x}_i)&=&-{V\over 2}\sum_n[u_n({\bf x}_i)v_n^*({\bf x}_i+{\bf \delta})
+v_n^*({\bf x}_i)u_n({\bf x}_i+{\bf \delta})]\tanh{E_n\over 2k_BT} ,
\label{DeltaV}
\end{eqnarray}
and the hole density is given by
\begin{eqnarray}
  p({\bf x}_i)=1-2\sum_n[|u_n({\bf x}_i)|^2f_n+|v_n({\bf x}_i)|^2(1-f_n)],
\end{eqnarray} 
where $f_n$ is the Fermi function.
Depending on the values of the potentials $V$ and $U$, it is possible
to have pairing amplitude with either $s$ or $d$ wave symmetry\cite{Franz,Ghosal,Ghosal2}. 

It has been shown\cite{Soininen}  that a superconducting
gap with $d$ wave symmetry calculated in a square lattice, can be written as
\begin{eqnarray}
\Delta_d({\bf x}_i)&=&{1\over 4}[\Delta_{\bf \hat x}({\bf x}_i)+\Delta_{-\bf\hat x}({\bf x}_i)
-\Delta_{\bf \hat y}({\bf x}_i)-\Delta_{-\bf \hat y}({\bf x}_i)].
\end{eqnarray}

Therefore we have used the above BdG theory to calculate the local
superconducting zero temperature doping depended $s$ and $d$ wave gap. In
Fig.(\ref{BdG}), we  show a typical set of results for $d$ wave as 
function of doping,  with  $V^{imp}=0$ and for a cluster
of $14\times 14$ sites. We have used
parameters which are appropriated to the HTSC, as we discussed in some of our
previous works\cite{Edson01,Edson02}: a hopping value of $t=0.35$eV,
next neighbor hopping $t_2=0.55t$, an on-site repulsion $U=1.3t$ and a
next neighbor attraction $V=-1.6$eV. Changing these parameters the
gap curve also changes but its qualitative form is not affected. This
calculation is to be used concomitantly with the phase separation
results from previous sections, since below $T_{ps}$, the system
has regions or islands of different doping levels. The consequences
of the BdG calculations, like those presented in Fig.(\ref{BdG}), will
be discussed in the next section in order to support the interpretation
of many physical properties associated with the HTSC.

\section{Discussion}

 As discussed in the introduction, it is very likely that phase separation 
is a fundamental process in the HTSC physics and therefore it must manifest itself
through  many experimental results.
In order to explore this fact, we have developed a 
formalism based on the CH differential equation 
which allows one to  quantitatively study 
the  HTSC phase  separation process. We take  the {\it upper pseudogap}
as the onset of phase separation because it starts 
in the underdoped region usually at very high temperatures 
($\approx800$K) where we expect neither Cooper pair formation 
nor fluctuation of these pairs and also because there are many arguments
against its identification with the superconducting gap\cite{Tallon}. 
In fact, the difficult to  associate
the experimental data at such high temperatures  with 
superconductivity led some authors to call it simply a 
crossover temperature\cite{TS,Emery}. Thus, assuming that the upper
pseudogap temperature line as that shown in Fig.(\ref{Tallonfig}) 
is the onset of phase separation, 
we have been able to provide  a simple interpretation
to the occurrence of a gap ($E_g$) at such high temperatures, to follow
the hole density time evolution and  how 
a small fluctuating (almost homogeneous) phase separates  into two main local 
densities ($p_-$ and $p_+$). Now we want to discuss some more
specific implications to the physics of HTSC if, in connection
with the above, we take the {\it 
lower pseudogap} as the local onset of superconductivity.

\begin{figure}[!ht]
\includegraphics[height=9cm]{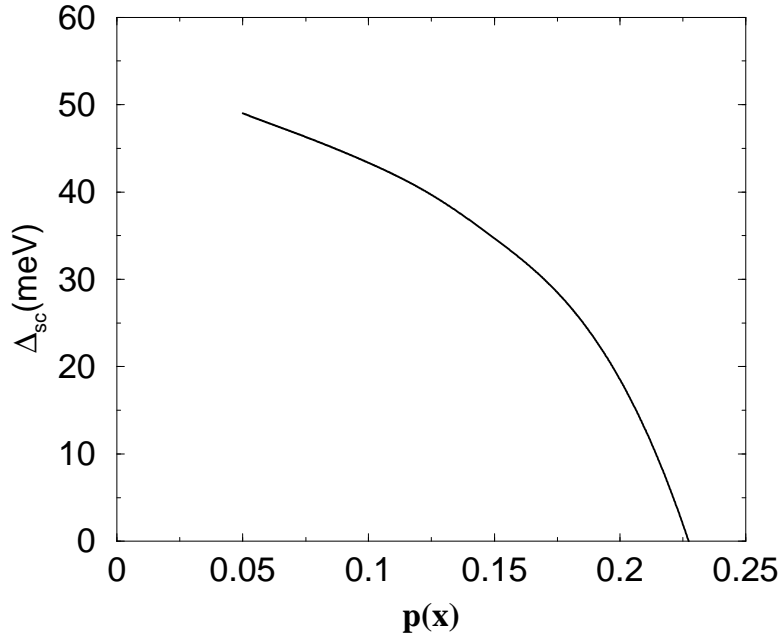}
\caption{ Results of the  calculation  for the zero temperature 
local superconducting gap using a $d$-wave BdG
superconducting theory. Since these calculations used a small
clusters, we can attribute the average density to a local density $p(\vec x)$.}
\label{BdG}
\end{figure}

The lower pseudogap has been attributed to the local mean field 
(MF) superconducting temperature or to the onset of pair
formation or superconducting fluctuation\cite{EK,Markiewicz,Uemura,Mihailovic}.
Starting in the underdoped region at temperatures usually near
the room temperature, it has been identified with the onset
of local superconductivity or with 
the appearance of small superconducting regions\cite{Mello03}.
This interpretation is supported by many different experiments,
the most directs being the Nernst effect\cite{Xu,Wang} and muon
spin rotation\cite{Uemura}. Following the theoretical predictions\cite{EK}
and the Nernst effect results\cite{Xu,Wang}, we assume that the lower
pseudogap vanishes at the strong overdoped region. 
Thus, in order to match the lower pseudogap, we have  calculated 
the zero temperature superconducting
gap $\Delta(p(\vec x))$ with a $d$-wave symmetry as described in
the previous section and  displayed in Fig.(\ref{BdG}). 
It is important to use  small clusters like
$8 \times 8, 12 \times 12$ and $14 \times 14$ in order to assure that
we are indeed calculating  the local properties but which are
larger than the coherence length.

It is interesting that the results of the local BdG 
zero temperature gap function   $\Delta(\vec x)$  have the same
qualitative form of the lower pseudogap\cite{TS,EK,Loram}, and it 
yields large values at low doping with its maximum near $p(\vec x)\approx 0.05$ and  
decreases continuously down to zero at the overdoped region. 
If the $\Delta(p)$ gap measured
in a compound with average hole density $p$ is assumed to be the corresponding
average value of all $\Delta(p(\vec x))$, we arrive that the 
$\Delta(p(\vec x))\times p(\vec x)$ is very similar to the
$\Delta(p)\times p$ curve. Indeed the  heat capacity
measurements (see  Fig.(8) of Tallon and Loram\cite{Tallon}) and the
ARPES (see Fig.(4) of  Harris et al\cite{Harris}) yield 
$\Delta(p)\times p$ curves with the
same qualitative form of the $\Delta(p(\vec x))\times p(\vec x)$
shown in Fig.(\ref{BdG}).

Thus, the lower pseudogap temperature,  which we denote $T^*(p)$, is the
onset of superconductivity and the superconducting regions grow as the
temperature is decreased below the $T^*(p)$,
but long range order is only possible at
the percolation limit among these regions, when
phase coherence is establish at $T_c(p)$. 
This scenario, with these two (phase-separation and local superconducting) pseudogaps,
is appropriate to interpret many non-convention HTSC features and
their main phase diagram, that is, the curves $T_{ps}(p)$ (the upper pseudogap
temperature),  $T^*(p)$ (the (superconducting) lower pseudogap
temperature) and $T_c(p)$, as  we
discuss below: 

We start with the discussion of
the many tunneling experiments results\cite{Renner,
Miyakawa,Suzuki,Krasnov00,Krasnov,Yurgens}: one of the most well known 
fact about these experiments is that they do not yield and special
signal at $T_c(p)$ and form a kind of "dip" that persists above $T_c(p)$ 
and dies off at $T^*(p)$. Our main point is that these experiments
are made over a finite region and 
always measure the average of all $\Delta(p(\vec x))$ in this region.
As the temperature is continuously raised from near zero, the
regions with weaker $\Delta(p(\vec x))$ (and lower $T_c(p(\vec x))$) 
become initially normal and, increasing more the temperature, many regions 
gradually turn from superconducting to
normal state but, all the local superconducting
regions are extinguished  only  at $T^*(p)$, not at $T_c(p)$. 
From the BdG calculations displayed in Fig.(\ref{BdG}),
we see that the regions with $p$ near $p_+$
yield gaps near the minimum value $\Delta(p_+)$ and we call it the lower
or weaker  branch. All the $\Delta(p(\vec x))$ in this branch which has
their local densities $p\le p(\vec x) \le p_+$,
vanishes before the temperature reaches $T_c(p)$ while those in the
strong branch $\Delta(p_-)$ with $p_-\le p(\vec x) \le p$ decreases
also continuously as temperature is raised but they are more robust and
totally vanishes only at $T^*$. These features are probed by  
tunelling experiments which, due to the very small mesoscopic
$p(\vec x)$ regions, usually measure the 
average of all these gaps. At low temperature, the average of many different gaps
are measured and as the
temperature is raised they all decrease and, firstly those in the weaker
$\Delta(p_+)$ branch and the ones in the second $\Delta(p_-)$ 
afterwards, vanish at different temperatures 
from zero, passing by $T_c(p)$ up to $T^*(p)$. Since  all different 
$\Delta(p(\vec x))$ vary continuously, there is not any special or 
different signal at $T_c(p)$. The  measured $dI/dV$ "dip" signal which came mostly
from the robust gaps in the  $\Delta(p_-)$ branch remains 
at temperatures well above $T_c(p)$ in the underdoped region,
decreases for compounds with increasing  doping level  $p$ since
$p_+$ and $p_-$ approaches one another, and in the overdoped
region  remains just for a few degrees above $T_c(p)$.
On the other hand, at the far overdoped region and above the critical
doping $p_c\approx 0.2$, there is no phase separation and the distribution
of $p(\vec x)$ is just a Gaussian-like distribution around the average hole doping $p$, 
consequently, the "dip" structure  remains only for a few degrees, proportional
to the distribution width.
This is well illustrated by Fig.(3) of Suzuki et al\cite{Suzuki} or
in the figures of Renner et al\cite{Renner}. 
As mentioned, these experiments see usually the average of many gaps in a 
given region, but, more recent refined experiments of Krasnov
et al\cite{Krasnov} and Yurgens et al \cite{Yurgens} were able to
distinguish between the gaps in the two average $\Delta(p_+)$ and $\Delta(p_-)$ 
branches. Fig.(1) of Krasnov et al\cite{Krasnov} shows clearly 
the weaker  $\Delta(p_+)$ (or superconducting
in their interpretation) peak fading away as
the temperature approaches $T_c(p)$ while the larger average $\Delta(p_-)$ (or
their pseudogap) "dip"
is almost unchanged. They have also shown  how applied magnetic fields up to 14T
destroy the weaker $\Delta(p_+)$  leaving again the stronger $\Delta(p_-)$ branch
untouched. Thus, the pseudogap signal remains after the pair coherence is lost 
because the isolated or local superconducting
regions left above $T_c$  are those with very large $\Delta(p(\vec x))$ 
or  $T_c(p(\vec x))\le T^*(p)$ and the 
temperature and fields to destroy the superconductivity in these regions
are much larger than $T_c(p)$ and the 14T used in the experiment\cite{Krasnov}. A
similar finding was provided by a STM experiment which measured a 
remaining pseudogap signal inside cores of Bi-2212 quantized vortices, 
where long range superconductivity is clearly destroyed\cite{Renner2}.
Furthermore, we predict that  the average $\Delta(p_-)$ maximum  
peak decreases slowly as the temperature  tends to $T_{ps}$
because $p_+$ and $p_-$ coalesce to $p$. This temperature
decreasing was recently measured and can also be seen in
Figs. (2 and 3) of Yurgens et al \cite{Yurgens}.

These results and our interpretation also agrees with the 
high magnetic field experiments which have measured simultaneously the closing of the 
pseudogap field ($H_{pg}$) and $T^*(p)$ by interlayer tunneling and
resistivity\cite{Lia}. Their reported results are for compounds
in the overdoped regime with $p\ge p_c\approx 0.2$, that is, for
compounds with doping level above the phase 
separation critical doping and $T^*(p)$ is just the maximum local  $T_c(p(\vec x))$
or lower pseudogap. At these doping
levels there is no phase separation and what Shibauchi et al\cite{Lia}
measured as $H_{pg}$ is the field that closes the  maximum 
{\it local} critical temperature which is $T^*(p)$ because
it is the largest of all locals $T_c(p(\vec x))$.
As they apply a magnetic field at low temperature, it destroys first the superconducting 
clusters with low  local $T_c(p(\vec x))$ and, as the field increases,
regions with larger values of the $T_c(p(\vec x))$ are destroyed. Increasing
even more the external field, eventually
it destroys the long range order or percolation among the superconducting
regions at the superconducting close field $H_{sc}$, leaving still some isolated 
regions which have larger $T_c(p(\vec x))$
than the phase coherence temperature $T_c(p)$.
Increasing more the field, one  reaches the closing field $H_{pg}=60$T which destroys
all the superconducting regions at $T^*(p=0.2)$. The closing field $H_{pg}$ must be very high
for compounds with lower doping, probably much higher than the 60T used by 
Shibauchi et al\cite{Lia} for a $p=0.2$ compound and that is the
reason why Krasnov et al\cite{Krasnov} did not see any
change in their optimally doped pseudogap dip at 14T. On the other hand,
a d-wave BCS with a Zeeman coupling yields good agreement with the data,
supporting the origin of the lower $T^*$ as the maximum local  superconducting
$T_c(p(\vec x))$\cite{Pieri}.
The fact that the local superconducting regions with large $T_c(p(\vec x))$
and low local doping (around $p_-$)
are very robust to an external magnetic field is also consistent with the Knight
shift measurements which have seen  the reductions 
of $1/T_1T$ and $K$ above $T_c$ from 
the values expected from the normal state at high temperatures 
in the overdoped region without any field effect up 
to 23.2T in the underdoped region\cite{Zheng}. 
Notice that, since the experiment of Shibauchi et al\cite{Lia} is performed 
with $p\ge 0.2$ samples, that is above the phase separation threshold $p_c$, 
therefore there is only one (Gaussian) dispersion of  local superconducting gaps
$\Delta(\vec x)$ and there is no gap $E_g$ associated with any
phase separation.

More recently, Hoffman et al\cite{Hoffman1,Hoffman2} and 
McElroy et al\cite{McElroy,McElroy2} developed a refined STM analyses
which let them to study the doping dependence and the electronic
structure of some compounds of the Bi-2212 family\cite{McElroy2}. 
They find a distribution of low
temperature local superconducting gap values $\Delta(\vec x)$ whose
average value $\Delta(p)$ and its width at half maximum increases
for compounds with average hole doping varying 
between $p \approx 0.19$ and $p \approx 0.11$.
The measured local values of $\Delta(\vec x)$ varies from
20meV to 70meV at regions with linear sides of approximate 55nm in length.
The low energy gaps exhibit periodic modulations consistent with
charge modulations like of a granular charge phase separation.
Their results, specially those shown in their Fig.(3A-3E), display a
distribution of mesoscopic scale regions local gaps of two types: \\
a) One type derived from a $dI/dV$
curve with sharp edges with values $<65$meV which they called coherence peaks. They 
interpreted this type of peaks as due to superconducting pairing
on the whole Fermi Surface arguing that this kind of  spectrum is consistent with
a d-wave superconducting gap\cite{Hoffman2}.\\
b) Another type of $dI/dV$  spectra display an ill defined  edges of a V-shape
gap with larger values than  +65meV what they
called zero temperature pseudogap spectrum. Furthermore, they
find that the a-type spectra are dominant for overdoped samples
($p=0.19$ and $0.18$) in which there is practically $0\%$ probability of occurring
spectra of b-type. The b-type spectra start to have a nonzero
probability for compounds with $p=0.14$ or below, and
for underdoped compounds like $p=0.11$ they find
more than $55\%$ of b-type spectra. It is not difficult
to explain these observations in terms of the CH phase
separation scenario: The
$p=0.19$ and $0.18$ compounds are near the phase separation threshold
and their $p(\vec x)$  distribution is essentially a Gaussian
type, $E_g$ is small and they measure a Gaussian distribution
of local superconducting gap values $\Delta(\vec x)$ or
a-type spectra. On the other limit, for the $p=0.11$ compound, according to the
phase separation histogram of Fig.(\ref{histoevol}), almost half
of the system has very low doping ($0<p(\vec x)<0.5$), the $p_-$
branch and almost half of the system is in the other $p_+$
branch ($0.17<p(\vec x)<0.22$). The regions with $p(\vec x)$ in the $p_+$ branch
exhibit a superconducting gap distribution of a-type spectra,
while regions in the  $p_-$ branch are mainly in the insulating region which
produces b-type spectra. These a- and b-type spectra are  mixed
in the intermediated $p=0.15$ and $p=0.13$ compounds and it is
in this near optimal doping region that both superconducting 
and the b-type gap have equally probability. 
Consequently, the gap maps found by Hoffman et al\cite{Hoffman1,Hoffman2} and
McElroy et al\cite{McElroy,McElroy2} are a clear manifestation
of the phase separation process in Bi2212.

There are many other experiments which we could discuss in the
light of the present phase separation theory but we believe that the
above discussion is sufficient to demonstrate that phase separation
process is central to understand many non-conventional HTSC
properties.

\section{Conclusions}

We have studied analytically  the problem of phase separation in HTSC
taking some current ideas on the possibility to identify the upper $T^*$ with
the onset of phase separation and the lower pseudogap as the onset
of d-wave superconductivity.
Our approach allows us to make quantitative calculations of
the phases separation process and to perform simulations which led
to granular and stripe patterns depending on the parameter and size
of the lattice, which are in agreement with current observations. Such
calculations might be also be pertinent to the physics of manganites.
It is also possible to get some insights on many general experimental results like:\\
i-The charge distribution becomes more inhomogeneous in the underdoped
region of the phase diagram where the stripes has been observed 
because $T_{ps}$ is larger in this region.\\
ii-The spatial variation or width of the local hole concentration $p(\vec x)$
increases as the temperature decreases.\\
iii-The fact that some materials exhibit granular  while others exhibit
stripe patterns may be related with the single crystal or palette size
of ceramic or granular samples. Our simulations indicate that larger lattices
favor stripe patterns while smaller ones favor granular patterns. \\
iv-The  spinodal decomposition reveals the importance of the sample 
preparation process, that is, samples with the 
same doping level may have different degree of inhomogeneity 
depending on the way they have been quenched through $T_{ps}$.
This would explain different results on the same kind of compounds
which has been very frequent in the HTSC.\\
v-The two different signal detected by refined tunelling experiments.\\
vi-The density os state  modulation measured by recent STM data.

In summary, the CH phase separation approach to HTSC 
in connection with local charge density dependent
BdG superconducting critical temperature calculation  is used to
explain the existence and nature of the two different pseudogaps and it 
provides novel interpretations on many non-conventional
features and inhomogeneous patterns. Therefore, our main point is that
we should regard the phase separation
process as one of the key ingredients of the HTSC physics.

\section{Acknowledgment}
We gratefully acknowledge partial financial aid from Brazilian
agencies CNPq and FAPERJ.

\end{document}